\documentclass[a4paper]{jpconf}
\usepackage{graphicx}
\usepackage{amssymb,amsfonts}
\usepackage{epsfig}
\usepackage{epstopdf}
\usepackage[all]{xy}
\usepackage{bm}
\usepackage{amsthm}

\newcommand{\be}{\begin{equation}}
\newcommand{\ee}{\end{equation}}
\newcommand{\bea}{\begin{eqnarray}}
\newcommand{\nn}{\nonumber}
\newcommand{\eea}{\end{eqnarray}} 

\newcommand{\gn}{\bar{\nabla}}

\def\inbar{\,\vrule height1.5ex width.4pt depth0pt}
\def\IR{\relax{\rm I\kern-.18em R}}
\def\IC{\relax\hbox{$\inbar\kern-.3em{\rm C}$}}

\begin{document}
\title{6$+$1 lessons from $f(R)$ gravity}

\author{Thomas P. Sotiriou}

\address{Center for Fundamental Physics, University of Maryland, College Park, MD 20742-4111, USA}

\ead{sotiriou@umd.edu}

\begin{abstract}
There has been a recent stimulus in the study of alternative theories of gravity lately, mostly triggered from combined motivation coming from cosmology/astrophysics and high energy physics. Among the proposed theories, one that has attracted much attention is $f(R)$ gravity. It is certainly debatable whether such a simplistic modification of General Relativity can constitute a viable alternative theory of gravitation. However, it is quite straightforward to see the merits of such a theory when viewed as a toy theory whose role is to help us understand the implications and difficulties of beyond-Einstein gravity. Under this perspective, I review some of the main lessons we seem to have learned from the study of $f(R)$ gravity in the recent past.
\end{abstract}

\section{Introduction}

Modifications of General Relativity (GR) by including higher order curvature invariants in the gravitational action have a long history. Indeed, it was just 1919 when 
Weyl, and 1922 when Eddington (the very man that three 
years earlier had provided the first experimental verification 
of GR 
by measuring light bending during a solar eclipse) started 
considering such terms in the action \cite{weyl,eddin}.  These early attempts were mostly driven by scientific curiosity and  a need to question and, therefore, understand the then newly proposed theory. Thus, they were quickly abandoned.  However, by the 1960's and 1970's new motivation coming from theoretical physics revived the field of higher-order gravity theories: Utiyama and De 
Witt 
showed that renormalization at one-loop demands that the 
Einstein--Hilbert action  be supplemented by higher order 
curvature terms \cite{Utiyama:1962sn}. Later on, Stelle showed 
that higher order actions are renormalizable (but not 
unitary) \cite{Stelle:1976gc} (see \cite{Schmidt:2006jt} for a historical review and a 
 list of references to early work). More recent results show that 
when quantum corrections or string theory are taken into 
account, the effective low energy gravitational action admits 
higher order curvature invariants \cite{quant1, quant2, quant3}. However, the 
relevance of such terms in the action was considered to be 
restricted to very strong gravity regimes and they were 
expected to be strongly suppressed by small couplings, as one 
would expect when simple effective field theory considerations 
are taken into account.

Recently, a new stimulus appeared in higher-order theories of gravity. This time the motivation came from cosmological and astrophysical observations. The latest datasets coming from different sources, 
such as the Cosmic Microwave Background Radiation (CMBR) and 
supernovae surveys, seem to indicate that the energy budget of 
the universe is  4\% ordinary baryonic matter, 
20\% {\em dark matter} and 76\% {\em dark energy} 
\cite{Spergel:2006hy,Riess:2004nr,Astier:2005qq,Eisenstein:2005su}. 
The term dark matter refers to an unkown form of matter, which 
has the clustering properties of ordinary matter but has not 
 yet been detected in the laboratory. The term dark energy is 
reserved for an unknown form of energy which not only has not 
been detected directly, but also does not cluster as ordinary matter 
does. The simplest model that fits the data adequately, is the so called concordance model or $\Lambda$CDM model (supplemented with some inflationary scenario), where a cosmological constant plays the role of dark energy and the nature of dark matter remains undetermined apart from the fact that it does not interact but gravitationally. However, even in the most optimistic scenario in which   the nature of dark matter will soon be determined in some accelerator, the known cosmological constant problems \cite{Weinberg:1988cp,Carroll:2000fy} would still plague the concordance model. One of the alternatives is to consider the unexpected observation as shortcoming of our macroscopic description of gravity itself: gravity should be by far the dominant interaction at cosmological scales so, could it be that a modification of GR could account for the unexplained observations?

It is tempting to combine the earlier motivation for corrections to GR coming from high energy physics with the more recent cosmological/astrophysical stimulus. This is indeed often done in the literature by proposing that some theory of gravity, which comes about as a low energy limit of a more fundamental theory and includes both ultraviolet and infrared corrections, could maybe account for all or some of the unexplained observations. However, there are two loopholes in such a suggestion: a) It is not clear exactly which one the fundamental theory should be or what is  its low energy limit, and b) the length scales at which we expect high-energy inspired corrections to GR to be important are orders of magnitude smaller than the length scales at which our puzzling observations have been made.

These two issues can be seen as caveats against modified gravity; or, in a more optimistic approach, they can be seen as opportunities: Problems, such as the ``unnatural" value of the cosmological constant have not found a solution within the realms of conventional (or even slightly unconventional) thinking. Therefore, they might as well be indications that something is wrong with the very core of this way of thinking. The large ambiguities  on how gravity works at small scales and which is the fundamental theory describing it seem to leave enough room for such speculations (indeed there are results already in the 
literature claiming that terms responsible for late time 
gravitational phenomenology might be predicted by some more 
fundamental theory, such as string theory \cite{Nojiri:2003rz}).

What seems to be the best way to proceed is probably the usual way when we are diving into the unknown in physics: consider a toy theory, probably of a very simple form, and use it as a tool to explore the limitations of the theory being questioned, in our case GR. This would allow us to test whether the very idea of modified gravity accounting for unexplained cosmological observations can be correct in principle, or what is the relation between the modifications required for that and those predicted by high energy physics. 

$f(R)$ theories of gravity seem to be very good candidates for such toy theories. They come about by straightforwardly
generalizing the Lagrangian in the Einstein--Hilbert 
lagrangian to be a general function $f$ of the Ricci scalar $R$, so that the action becomes
\begin{equation} \label{3}
S=\frac{1}{2\kappa} \int d^4 x \sqrt{-g} \, f(R),
\ee
where $\kappa\equiv 8\pi G$, $G$ is the gravitational constant, 
$g$ is the 
determinant of the metric and $R$ is the Ricci scalar 
($c=\hbar=1$). $f(R)$ theories of gravity make good toy theories for two reasons: a) They are sufficiently general  to 
encapsulate 
some of the basic characteristics of higher-order gravity, but 
at the same time they are simple enough to be easy to 
handle; b) there are unique among higher-order gravity 
theories, in the sense that they seem to be the only ones which 
can avoid the long known and fatal Ostrogradski instability  
\cite{Woodard:2006nt}.

Now, if we are to consider $f(R)$ gravity as a toy theory whose main scope was to help us learn something about beyond GR gravity, the obvious question to ask is what did we really learn so far? I attempt to answer this question here by going through some ``lessons" coming from $f(R)$ gravity (hence, the title). This talk is based on \cite{Sotiriou:2008rp}, where a longer discussion on the motivation of $f(R)$ gravity, as well as all on the other issues that are discussed here, is given (see also \cite{Sotiriou:2007}).

\section{What have we learned from $f(R)$ gravity?}

\subsection{Lesson 1: Choice of fundamental variables is crucial}

There are actually three versions of $f(R)$ gravity: Metric $f(R)$ gravity, Palatini $f(R)$ gravity, and metric-affine $f(R)$ gravity. In metric $f(R)$ gravity, the action 
\be
\label{metaction}
S_{met}=\frac{1}{2\kappa}\int d^4 x \sqrt{-g} \, f(R) 
+S_M(g_{\mu\nu},\psi),
\ee
where a matter action $S_M$ has been included and $\psi$ collectively denotes the matter fields, is varied with respect to the metric as usual. This yields the field equations
 \be
\label{metf}
 f'(R)R_{\mu\nu}-\frac{1}{2}f(R)g_{\mu\nu}- 
\left[\nabla_\mu\nabla_\nu -g_{\mu\nu}\Box\right] f'(R)= 
\kappa \,T_{\mu\nu}, 
\ee
where 
\begin{equation}
\label{set}
T_{\mu\nu}=\frac{-2}{\sqrt{-g}}\, \frac{\delta
S_M }{\delta g^{\mu\nu} }  ,
\end{equation}
a prime denotes differentiation with respect to the argument, 
$\nabla_\mu$ is the covariant derivative associated with the 
 Levi-Civita connection of the metric, and $\Box\equiv 
 \nabla^\mu\nabla_\mu$. Clearly, these equation are fourth order differential equations in the metric.
 
 Palatini $f(R)$ gravity comes about from the same action if we decide to treat the connection as an independent quantity. In this case, for the sake of clarity, we denote as ${\cal R}_{\mu\nu}$ the Ricci tensor constructed with this independent connection. ${\cal R}=g^{\mu\nu}{\cal R}_{\mu\nu}$ is the corresponding Ricci scalar and the action takes the form:
 \be
\label{palaction}
 S_{pal}=\frac{1}{2\kappa }\int d^4 x \sqrt{-g} \, f({\cal R}) 
+S_M(g_{\mu\nu}, \psi).
\ee
Independent variation with respect to the metric and the connection gives, after some manipulations, the following field equations
 \bea
\label{palf12}
f'({\cal R}) {\cal R}_{(\mu\nu)}-\frac{1}{2}f({\cal 
R})g_{\mu\nu}&=&\kappa \,G\, T_{\mu\nu},\\
 \label{palf22} 
\gn_\lambda\left(\sqrt{-g}f'({\cal R})g^{\mu\nu}\right)&=&0,
\eea
where $T_{\mu\nu}$ is defined in the usual way as in 
eq.~(\ref{set}) and  $\bar{\nabla}_\mu$ denotes the covariant 
derivative defined with the independent connection. Notice that when $f({\cal R})={\cal R}$, eq.~(\ref{palf22}) becomes the definition of the Levi-Civita connection and eq.~(\ref{palf12}) becomes Einstein's equation. So Palatini and metric variations both lead to GR for an action linear in the Ricci scalar. However, they lead to different theories for a more general action. In this sense, it is not straightforward which variation on should choose when generalizing the lagrangian to be a general function of the Ricci scalar.
Both metric and Palatini $f(R)$ gravity was first 
rigorously studied in \cite{Buchdahl:1983zz}.

In Palatini $f(R)$ gravity the connection, although independent, does not enter the matter action (this means that Palatini $f(R)$ gravity is a metric theory, according to the definition of \cite{willbook}). This implies that the covariant derivative of the matter fields which are usually present in the matter action are defined with the Levi-Civita connection (alternatively, the theory would include only very specific matter which does not couple to the connection, such as scalar fields, but this is not really an option for a gravity theory) \cite{Sotiriou:2006sr,Sotiriou:2006qn,Sotiriou:2006hs}. Additionally, the independent connection is really an auxiliary field: eq.~(\ref{palf22}) can be solved algebraically for the connection to give
\bea
\label{gammagmn}
\Gamma^\lambda_{\phantom{a}\mu\nu}&=&\frac{1}{f'({\cal 
R})}g^{\lambda\sigma}\Big[\partial_\mu \left(f'({\cal 
R})g_{\nu\sigma}\right)+\partial_\nu \left(f'({\cal 
R})g_{\mu\sigma}\right)\nn\\
&&-\partial_\sigma \left(f'({\cal 
R})g_{\mu\nu}\right)\Big].
\eea
At the same time, a contraction of eq.~(\ref{palf12}) with the metric gives
\be
\label{paltrace}
f'({\cal R}) {\cal R}-2f({\cal R})=\kappa\,T,
\ee
where $T=g^{\mu\nu} T_{\mu\nu}$. for a given $f$ this is an algebraic equation which can be solved to express ${\cal R}$ in terms of $T$.\footnote{See \cite{Ferraris:1992dx,Sotiriou:2008rp} for a discussion on the exceptional case where theis equation has no solutions or when $f({\cal R})\propto {\cal R}^2$.} Thus, using eqs.~(\ref{gammagmn}) and (\ref{paltrace}) one can algebraically express the independent connection in terms of the metric and the matter fields and eliminate it from the field equations. 

The way to give more substance to the connection is to restore its geometrical role and use it to define the covariant derivatives in the matter action, as proposed in \cite{Sotiriou:2006qn}. In this case the matter action would be of the form $S_M(g_{\mu\nu}, \Gamma^\lambda_{\phantom{a}\mu\nu}, \psi)$. Such theories are called metric-affine $f(R)$ theories. We refrain from giving their field equations here as we will not use them in what comes next. We refer the reader to \cite{Sotiriou:2006qn} for more details.

What should be clear after this discussion is that it is crucial to pick, not only the action, but also the fundamental fields which will be varied and assign to them a precise meaning. Remarkably, the distinction between Palatini and metric variation, which (at least at a classical level) was of little importance for the Einstein--Hilbert action, is critical for more general actions.

\subsection{Lesson 2: Equivalence with Brans--Dicke theory}

Metric $f(R)$ gravity appears as an extension of GR that includes no extra fields mediating gravity. Palatini $f(R)$ gravity appears to include a independent connection, so one could be trick to think that the underlying geometry is not pseudo-Riemannian. We already argued that the  independent 
connection is not really related to the geometry (at least as the latter is felt by the matter field which act as our probe). So appearances can deceive; and they do: both metric and Palatini $f(R)$ can be written as specific versions of Brans--Dicke theory \cite{Teyssandier:1983zz, Wands:1993uu, 
Barrow:1988xh,Barrow:1988xi,Flanagan:2003iw,Cecotti:1987sa, Wands:1993uu,Chiba:2003ir,Flanagan:2004bz,Sotiriou:2006hs}. Let us see how this is possible.

Starting from the action (\ref{metaction}) for metric $f(R)$ gravity, one can introduce a new field $\chi$ and write the dynamically 
equivalent action 
\be
\label{metactionH}
S_{met}=\frac{1}{ 2\kappa }\int d^4 x \sqrt{-g} 
\left[ f(\chi)+f'(\chi)(R-\chi)\right] 
S_M(g_{\mu\nu},\psi).
\ee
 Variation with respect to $\chi$ leads to the equation
 \be
 \label{1600}
 f''( \chi )(R-\chi)=0.
 \ee
 Therefore,
  $\chi=R$ if
$f''(\chi)\neq 0$, which reproduces the 
action~(\ref{metaction}).
Redefining the field $\chi$ by $\phi=f'(\chi)$ and setting
 \be
\label{defV}
V(\phi)=\chi(\phi)\phi-f(\chi(\phi)),
\ee
 the action takes the form
\be
\label{metactionH2}
S_{met}=\frac{1}{ 2\kappa }\int d^4 x \sqrt{-g} \left[ \phi 
R-V(\phi)\right] +S_M(g_{\mu\nu},\psi).
\ee
This is the Jordan frame representation of the  action of a 
Brans--Dicke theory with
Brans--Dicke parameter $\omega_0=0$ (also called ``O'Hanlon action'' as it was originally 
proposed by O'Hanlon in \cite{O'Hanlon:1972ya}).

We could follow exactly the same steps starting from action (\ref{palaction}) for Palatini $f(R)$ gravity. We would then arrive to the action
\be
\label{palactionH2}
S_{pal}=\frac{1}{2\kappa}\int d^4 x \sqrt{-g} \left[ \phi {\cal 
R}-V(\phi)\right] +S_M(g_{\mu\nu}, \psi).
\ee
This action is identical to (\ref{metactionH2}) apart from the fact that $R$ has been now replaced by ${\cal R}$. This is critical, as ${\cal R}$ is 
not the Ricci scalar of the metric $ g_{\mu\nu}$ and, therefore, action (\ref{palactionH2}) is not that of 
a Brans--Dicke theory with $\omega_0=0$. However, taking into account also the definition of $\phi$, we known that we can express the independent connection algebraically with respect to the metric  and $\phi$ using eq.~$\ref{gammagmn}$. Using this expression action (\ref{palactionH2}) takes the form
\be
\label{palactionH2d0}
S_{pal}=\frac{1}{2\kappa}\int d^4 x \sqrt{-g} \left(\phi 
 R+\frac{3}{2\phi}\partial_\mu \phi \partial^\mu 
\phi-V(\phi)\right)
+S_M(g_{\mu\nu}, \psi).
\ee
This is the action of a
Brans--Dicke theory with Brans--Dicke parameter $\omega_0=-3/2$. 

I will avoid to present the complete set of field equations for Brans--Dicke theory for the sake of brevity and give only the equation governing the dynamics of the scalar:
\be
(2\omega_0+3)\Box \phi+2 V-\phi V'=\kappa T.
\ee
This equation is enough to reveal that, it is not the $\omega_0=0$ case the one for which $\phi$ carries no dynamics as one might have expected judging from the absence of a kinetic term in the action, but actually the $\omega_0=-3/2$, which corresponds to Palatini $f(R)$ gravity.

The outcome of this discussion is twofold: Firstly, both metric and Palatini $f(R)$ gravity are equivalent to some version of Brans--Dicke theory (with all the implication this carries: for instance, this highlights the fact that Palatini $f(R)$ gravity is really a metric theory). Secondly, in metric $f(R)$ gravity there is only one extra scalar degree of freedom, where as in Palatini, the Brans--Dicke scalar is non-dynamical (so it is an auxiliary field, much like the independent connection). The moral is that not all representations of a theory are equally suitable for straightforwardly exhibiting some of the features of the theory (for a longer discussion see also \cite{Sotiriou:2007zu}).

\subsection{Lesson 3: Devil hides in the details}
During the last few years it has been understood that it is not that difficult to construct models within the framework of $f(R)$ gravity that lead to a desired cosmological background evolution (see \cite{Capozziello:2003tk,Carroll:2003wy,Nojiri:2003ft,Vollick:2003aw, 
Meng:2003bk,Meng:2004yf,Meng:2004wg, 
Allemandi:2004ca,Allemandi:2005qs, 
Sotiriou:2005hu, Sotiriou:2005cd} for some examples). Without getting into the details, the procedure is quite simple: Under the usual cosmological assumptions regarding symmetry and matter description, $f(R)$ gravity leads to modified Friedmann equations governing the dynamics of the universe. Choosing the function $f$ properly can lead to the desirable modification of the dynamics ({\em e.g.}~late time expansion). In fact, it has even been shown that, starting from a given scale factor $a(t)$ as a function of cosmic time,  one could  ``reconstruct'' the function $f$ (or at least a family of functions $f$) that leads to this evolution \cite{Capozziello:2005ku, Capozziello:2006dj, 
Nojiri:2006gh}.

However, the difficulties arise when one goes beyond background evolution and requires that some $f(R)$ can account also for the fine details in cosmological observations. For instance, the requirement that de Sitter spacetime (in models which do not admit Minkowski space as a maximally symmetric solution) be a stable solution gives important constraints \cite{Faraoni:2004bb, Faraoni:2005ie, Faraoni:2005vk}. Big Bang Nucleosynthesis is a delicate process that can lead to severe constraints as well ({\em e.g.}~\cite{Brookfield:2006mq}). The growth of cosmological perturbation and structure formation are other examples of processes that can rule out models and lead to strong viability constraints \cite{Bean:2006up,Koivisto:2005yc, Hwang:2001qk}. See \cite{Sotiriou:2008rp} for a more detailed discussion on cosmological constraints and a more complete list of references.

The lesson learned here is that the devil always hides in the details and proposing cosmologically viable $f(R)$ gravity models can be significantly trickier than it seems. The subtle details of the cosmological evolution are quite sensitive to the underlying gravity theory and as we are entering the era of precision cosmology it should become more and more difficult to deviate from GR without affecting the delicate picture of the standard model of cosmology.\footnote{Besides cosmology, another example where perturbations around the spacetime are much more revealing than the background itself, is black holes: even though the black hole solutions of GR are also solutions of $f(R)$ gravity model and one cannot distinguish between theories by mapping the spacetime around a black hole \cite{Psaltis:2007cw}, the perturbations around a black hole solution behave differently in different theories and, therefore, they allow for distinctions \cite{Barausse:2008xv}.}

\subsection{Lesson 4: Extra degrees of freedom are troublesome}

As mentioned in the Introduction, metric $f(R)$ gravity can avoid the Ostrogradski instability, unlike higher order gravity theories in general \cite{Woodard:2006nt}.  At the same time $f(R)$ theories have no ghosts (massive states of negative norm that cause apparent lack of unitarity), which is also not true for generic higher order theories of gravity \cite{quant3, 
quant1,  
Strominger:1984dn, Utiyama:1962sn,  Stelle:1976gc, 
Stelle:1978ww, 
Ferraris:1988zz}. Therefore, one could say that metric $f(R)$ gravity has been carefully chosen so that the extra degrees of freedom with respect to those of GR will not evidently lead to serious viability issues. 

Nevertheless, these criteria have proven inadequate and problems related with the presence of extra degrees of freedom are still plaguing metric $f(R)$ gravity. The most prominent is the instability initially found by Dolgov and Kawasaki \cite{Dolgov:2003px} for the model proposed in \cite{Carroll:2003wy} but can appear for more general functions $f$ \cite{Faraoni:2006sy}.\footnote{This instability does not appear in Palatini $f(R)$ gravity \cite{Sotiriou:2006sf}.} To see how this comes about let us  parametrize the deviations from Einstein gravity  as
\begin{equation}\label{1ter}
f(R)=R+ \epsilon \, \varphi(R) ,
\end{equation}
where $\epsilon $ is a small parameter with the dimensions of a 
mass squared and  $\varphi$ is arranged to be 
dimensionless. The trace of eq.~(\ref{metf}) yields
\begin{equation} \label{tracemet}
3\Box f'(R)+ f'(R) R -2f(R)=\kappa T ,
\end{equation}
and substituting eq.~(\ref{1ter}) gives
\begin{equation}\label{300}
\Box R +\frac{\varphi '''}{\varphi ''} \, \nabla^{\alpha} R \, 
\nabla_{\alpha} 
R+\frac{\left( \epsilon \varphi 
'-1\right) }{3\epsilon \, \varphi ''}\, R=\frac{\kappa \, T 
}{3\epsilon  \, \varphi''}\, +\, \frac{2\varphi}{3\varphi ''} .
\end{equation}
where it is assumed that  $\varphi '' \neq 0$. Consider now a small region of  spacetime in the weak-field regime 
and approximate {\em locally} the metric and the curvature by
\begin{equation}\label{4}
g_{\mu\nu}=\eta_{\mu\nu}+h_{\mu\nu} , 
\;\;\;\;\;\;\;\;\;\;\;\;\; 
R=-\kappa\, T +R_1 ,
\end{equation}
where $\eta_{\mu\nu} $ is the Minkowski metric and $\left| 
R_1/\kappa\, T\right| \ll 1$. Equation~(\ref{4}) yields, to first order in $R_1$,
\begin{eqnarray}
&& \ddot{R}_1 -\nabla^2 R_1 -\frac{2\kappa\, \varphi 
'''}{\varphi ''}\, \dot{T}\dot{R}_1+\, 
\frac{2\kappa \, \varphi '''}{\varphi ''}\, \vec{\nabla}T \cdot 
\vec{\nabla}R_1  \nn\\
&& +\frac{1}{3\varphi ''} \left( \frac{1}{\epsilon}-\varphi' 
\right) R_1=
\kappa \, \ddot{T}-\kappa \nabla^2 T -\, \frac{\left(\kappa\, 
T\varphi '+ 2 \varphi 
\right)}{3\varphi ''} , \nn\\\label{5}
\end{eqnarray}
where $\vec{\nabla}$ and $\nabla^2$ are the gradient and 
Laplacian  in Euclidean three-dimensional space, 
respectively, and an overdot denotes differentiation with respect 
to time. The function $\varphi$ and its derivatives are now 
evaluated at $R=-\kappa\, T$. The 
coefficient of $R_1$ in the fifth 
term on the  left hand side is the square of an effective mass 
and is dominated by the term  $\left( 
3\epsilon \,  \varphi '' \right)^{-1}$ due to the 
extremely small  value of  $\epsilon$ needed for these theories 
to reproduce the correct 
cosmological dynamics. Then, the scalar mode $R_1$ of the $f(R)$ 
theory  is stable if $\varphi '' =f''>0$, and unstable   if this 
effective mass is negative, {\em i.e.}, if  $\varphi''=f''<0$. 
The 
time scale for this instability  to manifest is estimated to be 
of the order of the inverse effective mass $\sim 10^{-26}$~s in 
the 
example $\epsilon \varphi(R)=-\mu^4/R$ 
\cite{Dolgov:2003px}. The small value  of $\varphi ''$ gives a 
large  effective mass and is responsible for the  small time 
scale over which the instability develops. 

This fatal instability for models with $f''<0$ manifests itself in the  linearized version of equation 
(\ref{tracemet}).  There are also recent claims that $R$ can be 
driven to infinity due to strong non-linear effects related to 
the same equation \cite{Tsujikawa:2007xu, 
Frolov:2008uf,Appleby:2008tv}. Both of these problems are strictly related to the extra scalar degree of freedom in metric $f(R)$ gravity and hint towards the following conclusion: no matter how carefully one constructs a theory with extra degrees of freedom, complications related to them can arise as extra degrees of freedom are very difficult to control.

\subsection{Lesson 5: Parametrized post-Newtonian expansion is difficult but fruitful}

Even though one might think that the parametrized post-Newtonian (PPN) expansion is more or less a standardized procedure, it actually turned out to hold a lot of surprises in $f(R)$ gravity. It took several papers and lengthy debates for this issue to be settled ({\em e.g.}~\cite{Barraco:2000dx,Dick:2003dw,Meng:2003sx,Chiba:2003ir,  Olmo:2005zr, Olmo:2005jd, Sotiriou:2005xe,Chiba:2006jp}, see also \cite{Sotiriou:2008rp} for a more detailed list of references). The main problems which were encountered and led to debates were: a) the equivalence with Brans--Dicke theory and whether this can be used for the PPN expansion, b) confusion related to the absence of flat spacetime as a solution in many models, c) misunderstandings regarding the difference between the existence of the Schwarzschild(-de Sitter) metric as a vacuum solution and a correct (post-)Newtonian limit, d) difficulty with non-analytic functions $f(R)$ and their expansion. Most of these problems  were not specific to $f(R)$ gravity and, therefore, solving them has led to a better understanding of the post-Newtonian expansion for alternative theories of gravity in general.

In spite of its difficulties, the post-Newtonian limit of $f(R)$ gravity has proved to be very fruitful after all: it led to strong viability constraints. Briefly summarized, the outcome of the relevant studies is the following: For metric $f(R)$ gravity, one can indeed use the equivalence with Brans--Dicke theory, but with some care ({\em e.g.}~the scalar is algebraically related to the curvature so it does not tend to settle to the minimum of the potential). This implies that, since the theory corresponds to a Brans--Dicke parameter $\omega_0=0$, it is ruled out (experiment require $\omega_0> 40 000$) unless the effective mass of the scalar is large. The effective mass is essentially the second derivative of the potential whose functional form depends on the form of $f$. It has also been shown that certain models of metric $f(R)$ gravity can exhibit ``chameleon'' behavior: the effective mass of the scalar depends on the curvature and, therefore, it can appear as large at small scales, making the mediated force short ranged and helping the theory avoid solar system constraints; however, the effective mass can become small at large scales, making the force long range and allowing the theory to have cosmological effects \cite{Cembranos:2005fi,Navarro:2006mw, Faulkner:2006ub, Starobinsky:2007hu}. In any case, the models have to be very carefully selected in order to be viable and at the same time lead to some interesting late time phenomenology (some examples do exist, {\em e.g.}~\cite{Starobinsky:2007hu,Nojiri:2007as})

The results for Palatini $f(R)$ gravity are much more intriguing. In this case, the known results from Brans--Dicke theory cannot be used because the $\omega_0=-3/2$ case is exceptional in the sense that the scalar is non-dynamical (the standard treatment assumes $\omega_0\neq -3/2$). In \cite{Olmo:2005zr, Olmo:2005jd} the PPN expansion for $\omega_0=-3/2$ Brans--Dicke theory and, consequently, Palatini $f(R)$ gravity was rigorously performed. Remarkably, as is pointed out there, and also in \cite{Sotiriou:2005xe} without the use of the equivalence between theories, whether a model will lead to the correct post-Newtonian behavior depends on the density. This had not been encountered before in alternative theories of gravity. Just to give a characteristic example, the zero-zero component of the PPN metric is given by
\be
\label{olmo1s}
h^{(1)}_{00} \left( t,\vec{x} \right)= \frac 
{2 G_{\rm eff} M_\odot}{r}+\frac{V_0}{6\phi_0}r^2+\Omega(T),
\ee
where $V_0$ and $\phi_0$ are the background values of the potential and the scalar respectively and $M_\odot \equiv \phi_0 \int d^3 \vec{x}' 
\rho \left(t, \vec{x}'\right)/\phi$. The
effective Newton constant $G_{\rm eff}$ and the post-Newtonian
parameter $\gamma$ are defined as
 \be
G_{\rm eff}
\equiv \frac{G}{\phi_0}\left(1+\frac{M_V}{M_\odot}\right),\quad
\gamma \equiv \frac{M_\odot-M_V}{M_\odot+M_V},
\ee
where $M_V\equiv \kappa^{-1} \phi_0 \int d^3 
\vec{x}'\left[V_0/\phi_0-V(\phi)/\phi\right]$. 
The striking term in eq.~(\ref{olmo1s}) is the last one: $\Omega(T)$ is some algebraic function of the trace of the stress-energy tensor $T$ whose functional form depends on the functional form of $f$. $T$ is in turn algebraically related to the density $\rho$ and so the PPN metric picks up an algebraic dependence on the density! This signals a serious shortcoming of Palatini $f(R)$ gravity, non-cumulativity, which will be discussed in the following section. Therefore, I refrain from saying more now.

To summarize, the PPN expansion for $f(R)$ gravity, even though it was a tedious procedure, deepened our understanding of post-Newtonian gravity in general and at the same time provided important viability constraints.

\subsection{Lesson 6: Non-cumulativity as the root of all evils!}

We just saw that in Palatini $f(R)$ gravity the PPN metric depends algebraically on the density. This issue turns out to be related with two other problems: a conflict between Palatini $f(R)$ gravity and the standard model of particle physics \cite{Flanagan:2003rb,Iglesias:2007nv} and the appearance of singularities on the surface of spherically symmetric matter configurations \cite{Barausse:2007pn,Barausse:2007ys,Barausse:2008nm}.

More precisely, it was shown in \cite{Flanagan:2003rb} assuming that matter is a Dirac field and in \cite{Iglesias:2007nv} assuming that matter is a scalar field that, when one goes to the local frame where spacetime is flat plus second order corrections, non-perturbative corrections and strong couplings appear in the matter action even at low energies. Since non-gravitational experiments are held at a local frame, these deviations from the standard matter actions contradict the results of such experiments. In \cite{Barausse:2007pn,Barausse:2007ys,Barausse:2008nm} on the other hand, spherically symmetric matter configuration where considered. It was shown that if one assumes a polytropic equation of state and attempts to match any interior solution to the unique Schwarzschild-de Sitter exterior, then singularities appear at the surface of the matter configuration when the polytropic index $\Gamma$ is in the range $3/2<\Gamma<2$. The presence of the singularities is essentially insensitive to the choice of the function $f(R)$ \cite{Barausse:2007pn,Barausse:2007ys}. Even though a polytrope is a very idealized form of matter, the importance of this results lies on two factors: a) There are physical matter configurations, such as a degenerate non-relativistic electron gas or an isentropic monoatomic gas, which can be described accurately by a $\Gamma=5/3$ polytrope; such configurations do not have a description in Palatini $f(R)$ gravity, rendering the theory incomplete. b) The polytrope is just used as an idealized approximation so that analytic results can be easily obtained, and the root of the problem lies deep within the structure of the theory. 

Lets us elaborate on this second point. This will also allow us to make the connection with the other problems discussed here. As mentioned earlier, the independent connection in Palatini $f(R)$ gravity is an auxiliary field, and one can express it algebraically in terms of the metric and the matter fields using eqs.~(\ref{gammagmn}) and (\ref{paltrace}) and eliminate it from the field equations. Indeed, if this is done, eqs.~(\ref{palf12}) and (\ref{palf22}) reduce to a single equation:
\bea
\label{eq:field}
G_{\mu \nu} &= &\frac{\kappa}{f'}T_{\mu \nu}- \frac{1}{2} 
g_{\mu \nu} \left({\cal R} - \frac{f}{f'} \right) + \frac{1}{f'} \left(
			\nabla_{\mu} \nabla_{\nu}
			- g_{\mu \nu} \Box
		\right) f'-\nn\\
& &- \frac{3}{2}\frac{1}{f'^2} \left[
			(\nabla_{\mu}f')(\nabla_{\nu}f')
			- \frac{1}{2}g_{\mu \nu} (\nabla f')^2
		\right], \nn
\eea
where $G_{\mu\nu}$ is the Einstein tensor of the metric $g_{\mu\nu}$. Note that, since ${\cal R}$ is algebraically related to $T$, $f({\cal R})$ and $f'({\cal R})$ are essentially function of $T$. $T$ contains usually first derivatives of the matter fields (or is algebraically related to the density for a perfect fluid). These imply that the right hand side (rhs) of eq.~(\ref{eq:field}) contains third derivatives of the matter fields (or second derivatives of the density). However, this equation is only second order in the metric. Therefore, the metric will have some algebraic dependence on the matter fields or the density! Gravity will not be cumulative, as in GR or most theories, and the characteristics of the gravitational field will depend on the local matter content.

This explains the behavior of the PPN limit. It also explains the conflict with particle physics (the terms on the rhs of eq.~(\ref{eq:field}) can be considered as coming from strong interactions). An easier way to think about this is to recall that in the equivalent Brans--Dicke theory the scalar is non-dynamical and algebraically related to the matter but non-minimally coupled to gravity. This leads to the strong couplings: when going to the local frame $\phi$ cannot be treated perturbatively. One has to solve for it in terms of the matter and replace back, introducing the new terms. {\em I.e.}~even though $\phi$ appears not to couple to matter, it enters the matter action though the coupling with the metric. 

The singularities found in \cite{Barausse:2007pn,Barausse:2007ys,Barausse:2008nm} are a manifestation of the non-cumulativity as well, in the sense that, in Palatini $f(R)$ gravity, the metric is bound to inherit any discontinuities in the matter and this would lead to a singularity. The simple example of polytropic matter configurations is a good demonstration of this, as  the range $3/2<\Gamma<2$ for the polytropic index is exactly the range for which the density on the surface is finite (and so there is a discontinuity since the density is zero in the exterior). However, matter is classically allowed to exhibit discontinuities. Remarkably, as it was mentioned, the singularities appear mathematically for essentially all functions $f$ (besides GR). There is only one case when one can question their physical significance: if they manifest themselves at scales for which the microstructure of matter has to be taken into account and the averaging procedure leading to a perfect fluid or a polytropic description is questionable. In this sense requiring the absence of such singularities provides a very strong constraint, ruling out all models with late time cosmological phenomenology, but even models with ultraviolet corrections \cite{Barausse:2007pn,Barausse:2007ys}. Only models with Planck suppressed correction seems to survive \cite{Olmo:2008pv}.

\section{Conclusions}

As is probably obvious by now, proposing and studying gravity theories other than GR is not an easy task. One can easily end up causing more problems that those the proposed modification of gravity is attempting to solve. Many of the shortcomings of $f(R)$ gravity become much more obvious after field redefinition and representation changes. The fact that it took some time to discover them seems to indicate that habit and affinity to specific formalisms and representation of a theory seem to have played a crucial role in the misunderstandings and the difficulties of $f(R)$ gravity. In any case, $f(R)$ gravity has been a very successful toy theory. It seems that we may have to go beyond simple  $f(R)$ models if we want to have a chance of explaining some of the current riddles of physics with a deviation from GR. However, our experience with $f(R)$ gravity  indicates that pursuing such an idea might be fruitful: possibly in providing a solution to some specific open problem, but most certainly in providing  insight into the most puzzling interaction, gravity. And this is an important lesson on it own.

\section*{Acknowledgements}
I am grateful to Valerio Faraoni, which was the coauthor of \cite{Sotiriou:2008rp} on which this talk was based, for numerous discussions on $f(R)$ gravity.
This work was supported by the National Science 
Foundation under grant PHYS-0601800.

\section*{References}
\bibliography{biblio}
\bibliographystyle{iopart-num}

\end{document}